\documentclass[usenatbib]{mn2e}

\usepackage{epsfig}

\newcommand{\bfig}{\begin{figure}}
\newcommand{\efig}{\end{figure}}
\newcommand{\btab}{\begin{table}}
\newcommand{\etab}{\end{table}}
\newcommand{\btabwide}{\begin{table*}}
\newcommand{\etabwide}{\end{table*}}
\newcommand{\figurewidth}{3.3in}

\newcommand{\lesssim}{{\la}}
\newcommand{\gtrsim}{{\ga}}

\begin{document}

 \title[The state of $Be^{7}$ in the Sun]{The State of ${\rm Be}^{7}$ in the Core of the Sun \\ and the Solar  Neutrino Flux}
     \author[N. J. Shaviv \& G. Shaviv]
   {Nir J. Shaviv$^{1,2}$, Giora Shaviv$^3$ \\
           $^1$ Racah Institute of Physics,
	 The Hebrew University, Giv'at Ram, 
	 Jerusalem, 91904, Israel\\
	 $^2$ Canadian Institute for Theoretical Astrophysics, 
     University of Toronto, 60 St.~George Street, 
     Toronto, ON M5S 3H8, Canada \\
     $^3$ Department of Physics  and Asher Space Research Institute,
     Israel Institute of Technology, 
     Haifa, Israel 32,000}
\date{Received March 3, 2002}
\pagerange{\pageref{firstpage}--\pageref{lastpage}}
\pubyear{2002}

\label{firstpage}
\maketitle

\begin{abstract}

The exact ionization state of ${\rm Be}^{7}$ in the solar core is
crucial for the precise prediction of the solar ${\rm B}^{8}$ neutrino
flux.  We therefore examine the effect of pressure ionization on the
ionization state of ${\rm Be}^{7}$ and all elements with $12 \ge Z \ge
4 $.  We show that under the conditions prevailing in the solar core,
one has to consider the effect of the nearest neighbor on the
electronic structure of a given ion.  To this goal, we first
solve the Schr\"odinger and then the Kohn-Sham equations for an ion
immersed in a dense plasma under conditions for which the mean
interparticle distance is smaller than the Debye radius.  The question
of which boundary conditions should be imposed on the wave function is
discussed, examined and found to be crucial.

Contrary to previous estimates showing that Beryllium is partially
ionized, we find that it is fully ionized.  As a consequence, the
predicted rate of the ${\rm Be}^{7}+e^{-}$ reaction is reduced by
20-30$\%$, depending on the exact solar model.  Since ${\rm Be}^{7}$
is a trace element, its total production is controlled by the
unchanged ${\rm He}^{4}+{\rm He}^{3}$ reaction rate, and its
destruction is determined by the rate of electron capture.  As the
latter rate decreases when the Beryllium is fully ionized (relative to
the case of partially ionized Be), the estimate for the abundance of
${\rm Be}^{7}$ increases and with it the ${\rm B}^{8}$ neutrino flux. 
The increase in $\phi_{\nu}({\rm B}^{8})$ is by about $20-30\%$.  The
neutrino flux due to ${\rm Be}^{7}$ electron capture remains
effectively unchanged because the change in the rate is compensated
for by a change in the abundance.  Hence the prediction for the ratio
of $\phi_{\nu}({\rm B}^{8})/ \phi_{\nu}({\rm Be}^{7})$ changes as
well.
\end{abstract}


\section{Introduction}

Classical calculations of solar models assume that all species are
fully ionized above ${\sim 10^{6}\rm K}$ (e.g.~\citealt{BP92}).  The main
reason probably being the saving of computer time, because detailed
ionization calculations are very CPU demanding.  Alternatively, one
can use tables for the equations of state which are calculated to high
accuracy.  However, one then has to interpolate for the relative
abundances which change continuously 
once diffusion takes place.  Indeed, from the point of view of the total gas pressure
and other thermodynamic quantities, the partial (or complete)
ionization of heavy species like ${\rm C, N, O, Mg }$ or even ${\rm Fe}$,
affect the number of free electrons at temperatures above a few
million degrees at a relative level of about $10^{-3}$, depending on
the exact mass fraction of the heavy elements.  Consequently, the
total pressure and speed of sound are affected at the same relative
level of accuracy.

Iben Kalata \& Schwartz (\citeyear{IKS67}, hereafter IKS67) examined the
ionization state of ${\rm Be}^{7}$ in the solar core, and concluded
that its K-shell electrons are partially bound (with a population
level of about 30\% depending on the exact location in the core). 
This fact significantly affects the predicted ${\rm B}^{8}$ neutrino
flux from the Sun.  The most important channel for the destruction of
${\rm Be}^{7}$ in the Sun is via electron capture, of which most are
free electrons.  However, if the ${\rm Be}^{7}$ ion has some bound
electrons then the rate of electron capture is enhanced, and with it
the generated ${\rm Be}^{7}$ electron capture neutrino flux (by about
20-30\% once averaged over the entire relevant region in the Sun). 
Thus, the exact occupation fraction of the ${\rm Be}^{7}$ K-shell is
important for the accurate prediction of the solar neutrino flux, and
the ratio $\phi_{\nu}({\rm B}^{8})/ \phi_{\nu}({\rm Be}^{7})$ in particular.  To
include the effects of the plasma, IKS67 assumed a Debye H\"uckel (DH)
potential and calculated the dependence of the ground state energy on
the environmental conditions.

The problem of obtaining the ionization state of Beryllium in the Sun
was later revisited by \citet{john92}.  The authors
analyzed the validity of the Debye-H\"uckel potential and found that 
the prerequisites for the validity are weakly violated. The authors claim  that
once the assumptions for the validity of the DH are strongly violated,  
``experiments show that the DH fails dramatically".  In particular, we
note the first point raised by the authors, namely the requirement to
have many particles in a Debye sphere needed for the validity of the
DH treatment.  This requirement implies that the interparticle
distance is significantly smaller than the Debye radius.  The authors
solve for the Beryllium atom assuming a DH potential using three
different methods (DH, self-consistent DH and Hartree) and find only
small differences in the ionization compared to IKS67.

\citet{Gruz97} discussed the ionization state of
Beryllium in the Sun assuming mean field screening, the density matrix
formulation, and a Monte Carlo method.  However, all within the
framework of a screened Coulomb potential.  The authors also discussed
the effects of fluctuations and found only minor effects.

If indeed Beryllium is partially ionized in the solar core, namely,
it keeps the K-shell electrons at least part of the time, then
several additional consequences follow.  These effects were hitherto
neglected in the prediction of the solar neutrino fluxes (predictions that 
assumed at the same time partial ionization of  ${\rm Be}^{7}$).

First, the screening of the nuclear reaction ${\rm Be}^{7}+p, $ which
is the competing ${\rm Be}^{7}$ destruction reaction, should be
calculated using the proper effective charge of the Be ion.  Currently
it is assumed to always have a charge of $+4e$.  This effect would
decrease the ${\rm B}^{8}$ neutrino flux.

Second, a similar correction to the screening should apply to the
higher Z reactions of CNO+p, affecting in this way the (small)
contribution of the CNO cycle to the total solar energy budget.  This
effect would suppress the CNO energy production because the effective
charge of the ion would be smaller and hence the electrostatic
screening energy would be smaller as well.

Third, the exact point at which various ions become completely ionized
affects the entropy density in the outer convective zone of the sun,
and with it the solar structure.

As the accurate prediction of the solar neutrino flux is so important,
the purpose of this contribution is to re-examine the ionization state
of the heavy species in the solar core, and in particular the
ionization state of the trace element ${\rm Be}^{7}$.

The question to which extent does the ${\rm Be}^{7}$ or any other
heavy ion, retain its K-shell electrons is usually analyzed in two
steps (IKS67).  The first step is to apply the simple Saha equation
assuming that the structure of the ${\rm Be}^{7}$ atomic energy levels
is unaffected by the dense plasma.  The second step is to account for
the effect of the plasma on the energy levels of the ${\rm Be}^{7}$ by
assuming a smooth DH potential and calculating the energy levels under
the DH potential.  Once the new energy levels are known, the electron
population in the levels can be re-evaluated using a Saha equation
which incorporates the revised energy levels.  This approach is
justified only as long as the plasma effects are small perturbations.

As we shall show, the conditions in the core of the Sun are very
peculiar and the number of particles inside a Debye sphere, 
$N_D \approx$~{\em few} and {\it not} very large compared
to unity.  Hence, the necessary condition for the validity of the
Debye H\"uckel theory is not satisfied.  Moreover, the conditions in
the core of the Sun are such that the mean distance between ions in
general, $\langle r_{s} \rangle = (4 \pi n/3)^{-1/3}$, and between a
proton or an $\alpha$-particle and a Beryllium ion in particular, is
of the order of $2R_{B}(Z=4) $ (the Bohr radius in a nucleus with
charge $Z=4$) and hence the picture of an ion with an electronic shell
inside a Debye-H\"uckel potential, is not strictly valid.  Here, $n$ is the
 number density of ions while the index
$s$ in $\langle r_{s} \rangle$ corresponds to a calculation employing
spherical packing of ions. When the DH
theory applies, it means that there are many ions inside a Debye
radius and the mean distance between ions is much smaller than the Debye
radius.  The electronic structure of the ions is  then affected first
and foremost by the close ion rather than by the Debye cloud and its
large radius.   This point, which is essential in this
paper, will be discussed at length since this situation dictates a
different boundary condition which subsequently leads to different
energy levels and a different ionization state (under the same 
thermodynamic conditions).

The paper is structured as follows.  We first question and analyze the
effective potential to be used under the conditions prevailing in the
solar core.  Then we repeat the two steps analysis of IKS67.  We next
proceed to examine the pressure ionization at $T=0$ assuming a Coulomb
potential.  In view of the doubtful validity of the DH potential under
the conditions prevailing in the core of the Sun, we repeat the
calculation assuming the Schr\"odinger and later the Kohn-Sham
equations.  We find that ${\rm Be}^{7}$ is fully ionized at a lower
density (and temperature) then previously calculated.  Finally, we
examine the effect of the complete ionization of ${\rm Be}^{7}$ on the
predicted solar neutrino flux according to different sets of nuclear
reaction cross sections.

\section{Which potential?}
\label{sec:whatp}
The classical calculation of the atomic energy levels and pressure
ionization assumes a smooth (in time and space) potential.  We first
examine the assumption that any applied potential can be assumed to be
smooth for the specific problem of the  structure of the electronic level
of a given ion under the conditions prevailing in the core of the Sun.

For the assumption that the potential is smooth to be valid, the
fluctuations of the plasma must be much faster than the motion of the
electron in the bound orbit such that the average smooth value can be
taken (for the calculation of the bound state) rather than the
instantaneous one.  We assume at the beginning that the plasma does
not affect the energies of the bound levels.  (The effect of the
plasma is to make the energy levels shallower so the electrons would
move even slower.  Thus, the argument presented here is a
conservative one.)

The Bohr radius in a Hydrogen-like  ion with a charge Z (in vacuum) is given by:
\begin{equation}
	a_{0}={h^{2} \over 4 \pi^{2} m_{e} e^{2} Z}={ 1 \over Z}0.528 \times 
	10^{-8}{\rm cm}.
\end{equation}¥
 The classical velocity is:
\begin{equation}
	v=2 Z e^{2} \left({\pi  \over \hbar } \right)^{1/2},
\end{equation}
and the period is 
\begin{equation}
	P={h^{3}  \over 4 \pi^{2}m_{e} e^{4}Z}.
\end{equation}
There are two sources for fluctuations in the potential, those
caused by the protons and those caused by the electrons.  The typical
time scales of the fluctuations are
\begin{eqnarray}
	\tau_{e} = {\langle r_{s} \rangle \over v_{th} (ZN_{D})^{1/2}}=
	{\langle r_{s} \rangle m_{e}^{1/2} \over (3kT)^{1/2} (ZN_{D})^{1/2}} \\
	\tau_{p} = {\langle r_{s} \rangle \over v_{th} N_{D}^{1/2} }=
	{\langle r_{s} \rangle m_{p}^{1/2} \over 
	(3kT)^{1/2} N_{D}^{1/2}},
\end{eqnarray}
where $\langle r_{s} \rangle=1/(4 \pi n_{{\rm ion}}/3)^{1/3}$ is the mean
interparticle distance and $n_{{\rm ion}}$ is the number density of ions. 
$v_{th}$ is the relevant thermal velocity.  The number of particles in
the Debye sphere is given by:
\begin{equation}
	N_{D} = {4 \pi \over 3} R_{D}^{3},
\end{equation}
where 
\begin{equation}
	R_{D}= \sqrt{{ kT \over 4 \pi e^{2} \sum_{j}(Z_{j}^{2}+Z_{j}) 
	n_{j}}}
\end{equation}
is the Debye radius for this mixture.  $n_{j}$ is the number density
of specie $j$ with charge $Z_{j}$.  The above expression for the Debye
radius assumes that both the electrons and ions contribute to the
supposedly DH potential.  For simplicity we assume $N_{D}^{1/2}
\approx 3$ and obtain that for $n=10^{26}\# /{\rm cm}^3$, ${\rm
T}=1.5\times 10^{7}$K and a pure Hydrogen plasma,
\begin{equation}
	{P \over \tau_{e} }\approx 0.5 ~~~~{\rm and} ~~~~{P \over \tau_{p} 
	}\approx  0.01.
\end{equation}
Note that when some of the ions are Helium ions (in the core of the
present Sun about half the ions are He), $N_{D}$ decreases even more. 
What is the implication of this result?  The fluctuations due to the
electrons are of about the same timescale as the period of the
electrons in the K-shell in a single ion in vacuum, while the protons
in the plasma have a much longer time scale.  Hence, it is not
justified to treat the contribution of the electrons to the DH
potential (felt by the electron) as a smooth potential in time.  On
the other hand, since there are only few protons in the
Debye radius their contribution to the potential is smooth in time but not in space
and certainly not spherical.  Additional arguments that question the
validity of the potential are given by \citet{john92}.  In what
follows, we assume that all potentials are temporally smooth and
spatially spherical.

\section{The ionization of Beryllium in the Sun}
\subsection{The state of ionization ignoring plasma effects}
Next, we discuss the ionization state of ${\rm Be}^{7}$ in the core of
the Sun - the classical way.  If one adopts the Saha equation,
ignoring screening and the excited energy levels (thus including only
the ground states in the partition functions), then the probabilities
$f_{1}$ and $f_{2}$ that one or two K-shell electrons are associated
with any given ${\rm Be}^{7}$ nucleus are given by (IKS67)
\begin{eqnarray}
	f_{1}&=& { \eta \over 1 + \eta + 0.25 {\eta}^{2} \exp ({-\delta 
	\chi/kT})}, \\ \nonumber
	f_{2} &=&  0.25 {\eta} \exp ({-\delta \chi/kT})f_{1},
\end{eqnarray}
where     
\begin{equation}
	\eta  = n_{e}({h^{2}  \over 2 \pi m k T})^{3/2}\exp (\chi_{1}/kT).
\end{equation}

 Here $\chi_{1}=216.6 {\rm ~eV}$ is the forth ionization potential of
 the ${\rm Be}^{7}$ atom, $\chi_{2}=153.1 {\rm ~eV}$ is the third
 ionization potential of the ${\rm Be}^{7}$ atom, and $\Delta \chi =
 \chi_{1} - \chi_{2} = 63.5 {\rm ~eV}$.  These values correspond to
 the limit of vanishing plasma density.  $n_{e}$ is the number density
 of the free electrons, most of which are contributed by Hydrogen and
 Helium and are independent of the state of trace elements like
 Beryllium.  Thus, $n_{e}$ can be treated as fixed.

The application of the Saha equation in the above form to the core of
the Sun ($\rho =158 {\rm ~g~cm^{-3}}$, ${\rm T} =1.57 \times
10^{7}{\rm K}$, $X=0.36$ \& $Z=0.02$) yields $f_{1}=0.320$ and
$f_{2}=0.038$, implying that the ${\rm Be}^{7}$ keeps its last
electron for about a third of the time.

As the relevant ions are in a plasma, the traditional 
procedure to correct for the plasma effect is to replace the pure
Coulomb potential with a DH one \citep{RGH70}. 
This is for example the procedure IKS67
evaluated the plasma corrections for the energy levels of ${\rm
Be}^{7}$ in the solar core.

The Saha equation in the above form ignores electron degeneracy,
exchange effects and pressure ionization.  The electron degeneracy
introduces a small correction under the conditions prevailing in the
Sun.  As we will shortly demonstrate, exchange and pressure ionization
are significantly more important.  In what follows we do assume in
spite of the previous reservations, a smooth static DH potential
contributed by the electrons and ions.  Moreover, we assume it to be
relevant in a statistical sense only.

\subsection{Screened potential: taking into account the plasma effects}

After performing the above estimate, IKS67 turned to evaluate the
ground state of the $Z=4$ ion assuming a smooth DH screened potential
in which both the protons and the electrons are taken into account. 
We ignored the questions raised in the previous section concerning the
validity of the potential for our particular purpose here (ionization
in the core of the Sun), repeated their calculation and confirmed
their results with respect to the Hydrogen like ion with $Z=4$.
\citet{RGH70} calculated the bound states of static screened coulomb
potential and formulated their results in terms of the screening
length.  We also repeated their results for the $1s$ state and the results
are shown in fig.~\ref{fig:screenE}.  Clearly, as the Debye length
approaches the Bohr radius of ions with charge Z, there are no more
bound states.  The boundary condition on the wave function in this
case is $\psi(r \rightarrow \infty ) =0$.

\bfig 
\epsfig{file=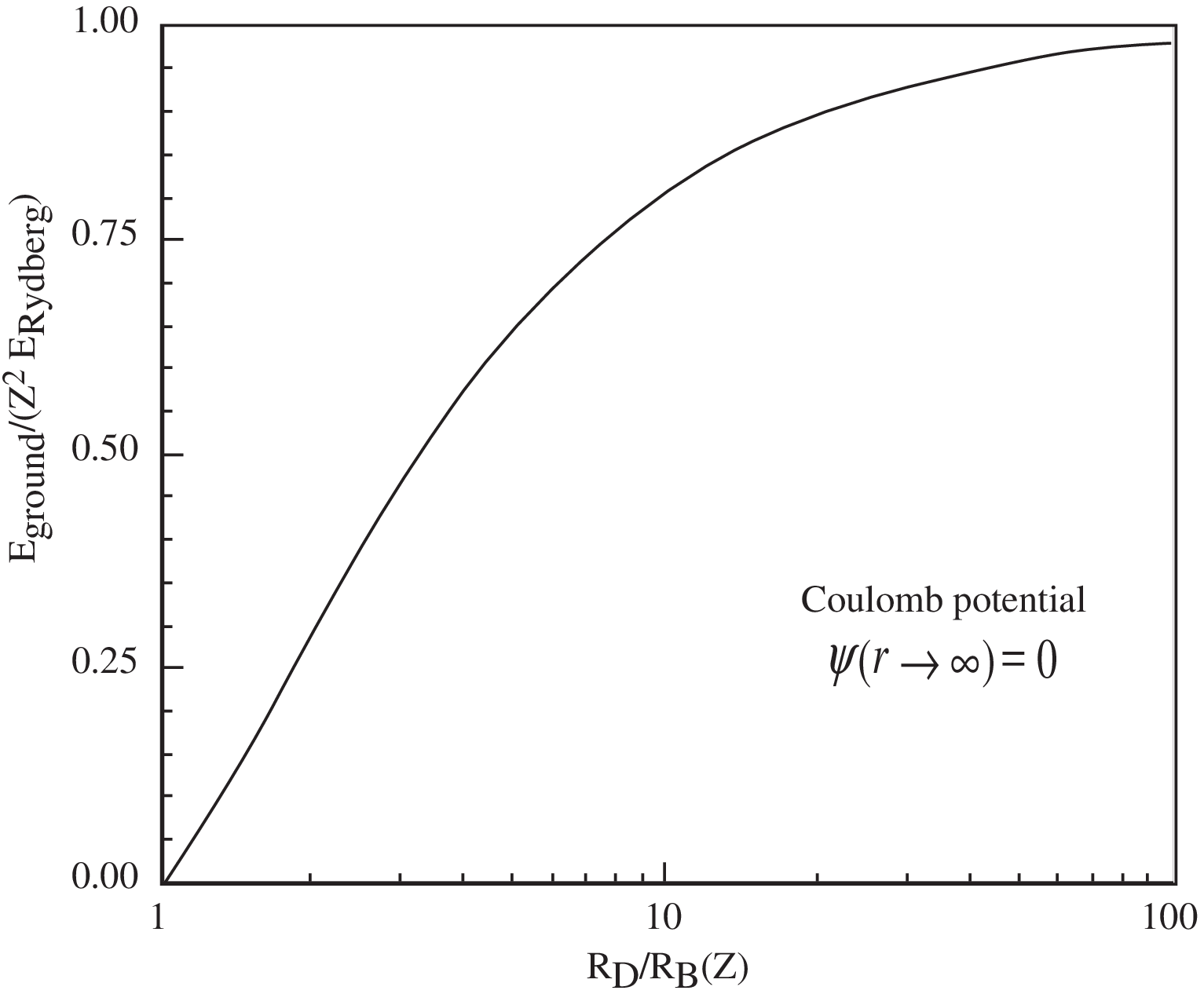,width=\figurewidth} 
\caption{The
energy of the ground state for a screened Coulomb potential as a
function of the screening length.  Note the units of the energy
$(Z^{2} E_{\rm Rydberg})$ and Debye length scale $(R_{D}/R_{B}(Z))$ where
$R_{B}(Z)$ is the Bohr radius of an ion with charge Z. }
\label{fig:screenE}
\efig

The calculation of the plasma effects on the triply ionized 
${\rm Be}^{7}$ ion is more complicated because of the partial screening of
the nucleus by the bound electron.  To overcome this problem, we used
the following approximate method.  We looked for the eigenvalue in the
low density limit and searched for the effective charge that will
reproduce the measured ionization potential of 153.1 eV. We found that
this charge is $Z=3.3544$.  We then repeated the calculation of the
plasma effect on the bound states assuming this charge.  The results
are shown in fig.~\ref{fig:hi12-diff}.  We then used the new values
for the ionization potential in the Saha equation to find the revised
$f_{1}$ \& $f_{2}$.  The comparison between the values used with and
without the plasma correction are shown in fig.~\ref{fig:Be7-sun}, where
the actual run of the occupation numbers in the Sun is given.  It is surprising
that the differences in the ionization come out to be quite small.

\bfig 
\epsfig{file=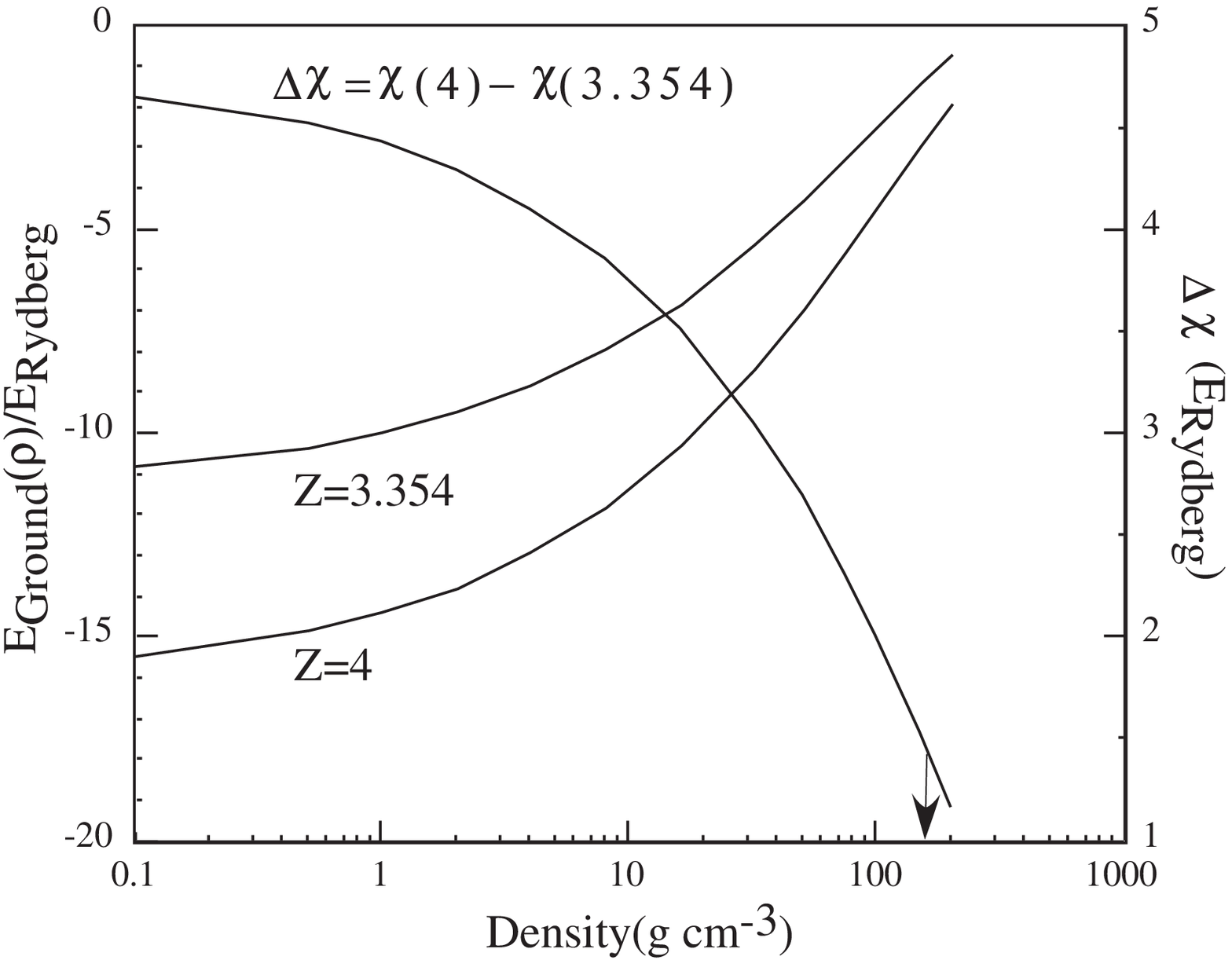,width=\figurewidth} 
\caption{The
run of the ground energy levels of a Be ion with one and two electrons 
as a function of 
density for a Debye H\"uckel potential and
$\psi(r\rightarrow \infty)=0$ as the boundary condition.  The temperature,
which enters via the Debye radius, is taken as constant at $1.57\times
10^{7}{\rm K}$.  Note that the temperature in the Sun decreases with
density and hence the temperature and with it the Debye radius are
overestimated here for densities lower than $150 {\rm ~g~cm^{-3}}$. 
The correction for the accurate temperature is small.  The purpose of
the figure is to show the conditions in the solar core.  The
small arrow marks the density at the core.}
\label{fig:hi12-diff}
\efig

\bfig
\epsfig{file=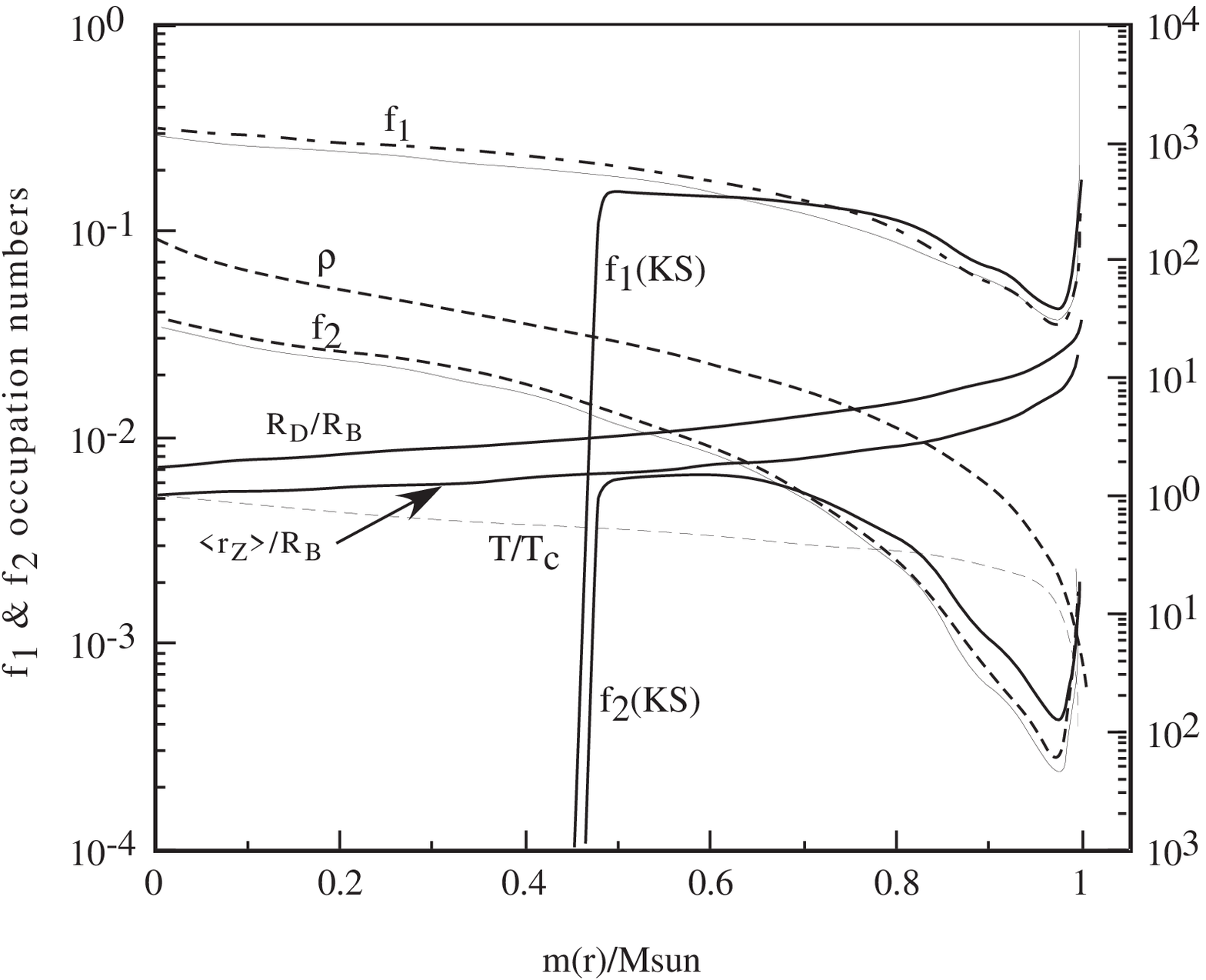,width=\figurewidth}
\caption{The occupation numbers $f_{1}$ and $f_{2}$ of Be throughout
the Sun as a function of the solar mass fraction.  The broken lines
are the results of IKS67.  The continuous lines are the present
results after incorporating the effective charge of the ${\rm Be}^{3+}$ ion
($Z_{\rm eff}=3.354$).  The curves marked with $f_{1}(KS)$ and
$f_{2}(KS)$ are the results assuming $\psi '=0$ on the cell boundary. 
These results are the actual run of the occupation numbers in the Sun. 
Also shown are the density and temperature (in units of the central
temperature) in the present day Sun.  The run of $R_{D}/R_{B}$ and
$\langle r_{Z} \rangle /R_{B}$ 
are shown as well.  $T/T_c$, $\rho$, $R_D/R_B$ and $\langle r_{Z} \rangle /R_{B}$
are all shown on the left axis. The occupation numbers are shown on the right axis.}
\label{fig:Be7-sun}
\efig

The particular results for the binding energy (calculated for a DH
potential) as a function of the density are shown again in figure
\ref{fig:Be-Debye} along with the run of the ratios $R_{D}/R_{B}(Z=4)$
and $\langle r_{s} \rangle /R_{B}(Z=4)$.  The Debye radius and the
mean interparticle distance are calculated assuming $X=0.34, Y=0.68$
and $Z=0.02$, a composition which is close to the one at the solar core
today.

\bfig
\epsfig{file=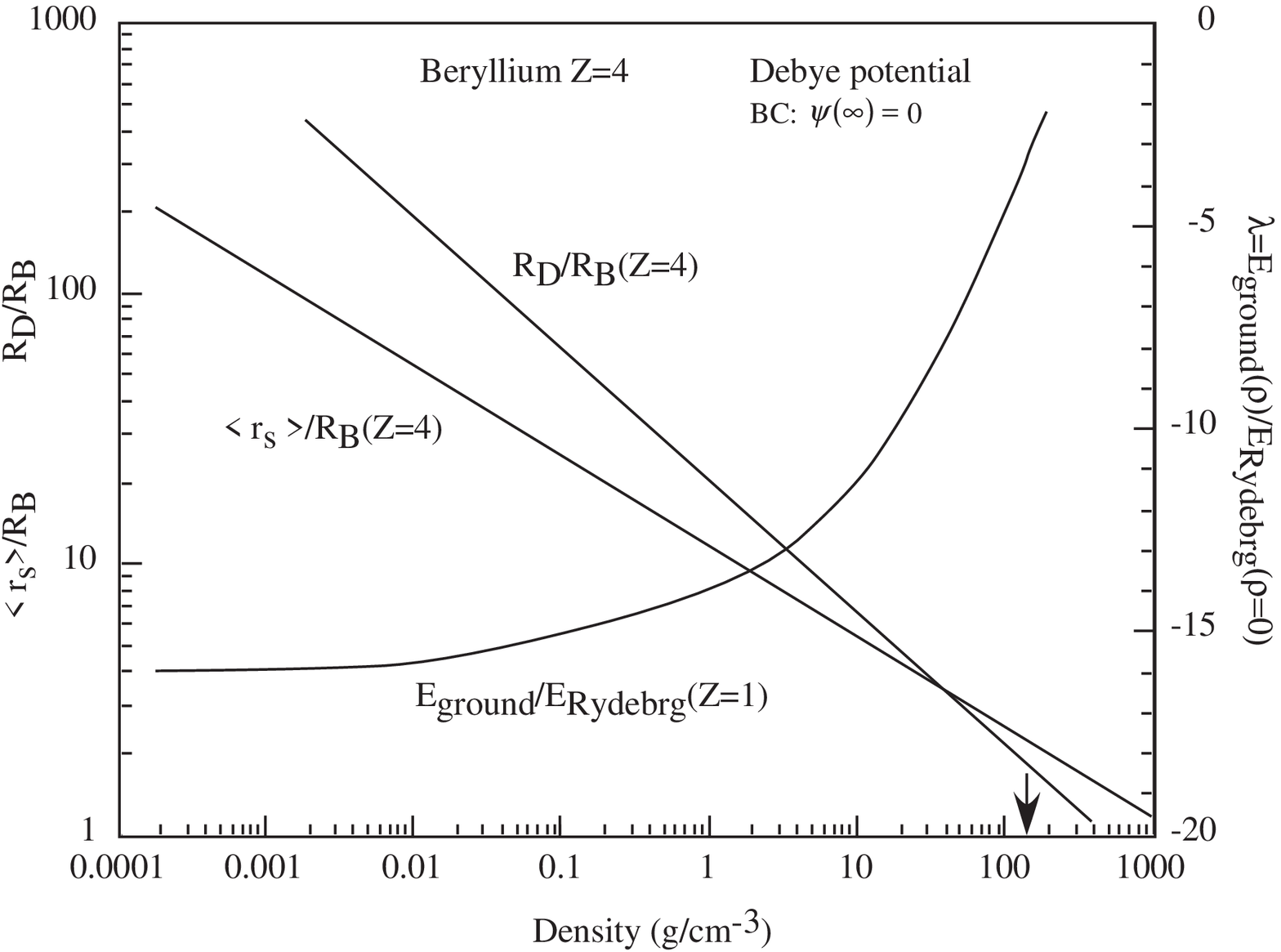,width=\figurewidth}
\caption{The run of the energy of the ground state as a function of 
density assuming fixed temperature and composition. Also shown are 
$R_{D}/R_{B}(Z=4)$ and $\langle r_{s} \rangle /R_{B}(Z=4)$. The arrow 
marks the density in the center of the Sun. }
\label{fig:Be-Debye}
\efig

We notice that when the density approaches the density in the solar
core, namely about $150~{\rm g~cm^{-3}}$, (a) the Debye radius becomes
of the order of the mean interparticle distance and hence the
approximation of a smooth Debye screened potential loses its validity,
emphasizing once more the conclusion reached in section
\ref{sec:whatp}.  (b) In the solar core, we find that 
$R_{D} \approx R_{B}(Z=4)$ and therefore the
probability for complete ionization of the ${\rm Be}$ is very high.

However, the more important question here is the ratio of the mean
interparticle distance to the Bohr radius since we are interested in
the possibility that the ions of Beryllium still have bound electrons. 
Fig.~\ref{fig:Be-Debye}, which depicts a graph for the 
value of $\langle r_{s} \rangle $, indicates that this value at
the center of the Sun it is close to $R_{B}(Z=4)$. Therefore, it is a
delicate question whether the Beryllium ions possess any bound
electrons.  Finally, we point out that $\langle r_{s} \rangle$, which 
is depicted in the figure, is the mean interparticle distance irrespective
of their type.  As we shall show, it is an underestimate in the case
of a Beryllium ion embedded in Hydrogen and Helium ions.

\section{Pressure ionization at $T=0$}
\subsection{The boundary conditions}
As discussed above, the classical method to evaluate the degree of
ionization in a stellar plasma with a finite temperature is first to
assume given mean distances between the particles, and assume that
they are at rest, namely that $T=0$.  (However, we do keep the finite
temperature in the calculation of the Debye radius).  Once the energy
levels are known, the effect of the temperature via the Boltzmann
relation (leading to the Saha equation) is taken into account.  In an
actual plasma, the distance between the particles has a distribution
and hence there is a distribution of cases.  One assumes that the
average of the results for the distribution is equal to the result for
the average.  We turn now to the $T=0$ case.

The question of pressure ionization is discussed  by \citet{Chiu99}
within a general discussion about the energy levels of atoms in plasma
and follows \citet{Rous74} and \citet{RGH70}, where the
effect of the plasma was simulated by a screened potential.  Pressure ionization
depends primarily on the density. However, when the
relevant scale is the Debye length, some effect of the temperature on
the energy levels enters through the back door via the dependence of
the Debye screening length on the temperature, which as stated before
we do keep finite in the calculation of the potential (in the case
that a DH potential in assumed).

We distinguish between two possible situations:

Case A: $R_{D} \gtrsim \langle r_{s} \rangle  ~ {\rm or} ~N_D\gtrsim 1$ and 

Case B: $R_{D} \lesssim \langle r_{s} \rangle  ~ {\rm or} ~N_D \lesssim 1$. 

The physical  difference between the two cases 
is expressed in the boundary conditions imposed on the wave function
 in the problem of the electronic structure of the ion. 
In the first case, the nearest neighbor is closer than the Debye distance 
and hence one expects it to affect the electronic structure of the ion much
 more than the fact that the two ions are inside the same potential well.  
In the second case, the nearest neighbor is
further away than the Debye radius and hence the boundary condition on
the wave function with a DH potential can be $\psi(r\rightarrow \infty
)=0$.  On the other hand, in the first case, the boundary condition
should take into account the close ion.  As the speed of the
perturbing ions is of the same order as the speed of the ion under
consideration, the effect of the ions inside the Debye sphere cannot
be averaged into a mean potential.  The basic requirement to consider
the effect of the nearest neighbor directly (and not via a smooth
potential averaged over many ions) leads to the idea of a Wigner-Seitz
unit cell or ion-sphere.

We can look also on the problem in the following way: If the bound state 
electronic wave function of a given ion overlaps significantly the bound
electronic  wave function of nearby ions the electron cannot be considered
 as bound (cf. \citealt{MW98}).

We approached the problem of pressure ionization at $T=0$ in two steps
which represent successive approximations.  In the first step we solve
the Schr\"{o}dinger equation for the ${\rm Be}^{7}$ ion under the
assumption that there is another nucleus at a distance $\langle r_{s}
\rangle$ away.  In our particular case, the plasma
contains ions with different charges and one cannot state that the
ion sphere  of all ions is identical.  Hence the boundary condition must be imposed 
 at $ \langle r_{z} \rangle=  \alpha \langle r_{s} \rangle$, 
where $\alpha$ is soon to be determined.    In the pure periodic
case one should apply the Bloch condition (\citealt{Mard00}, see also
\citealt{Lai91}).  We assume for simplicity spherical symmetry and hence
the Bloch condition becomes the requirement that
\begin{equation}
\left.{d \psi \over dr}\right|_{r=\alpha \langle r_{s} \rangle}=0.
\label{eq:bloch-cond}
\end{equation}
The coefficient $\alpha$ is determined from the
condition that the force vanishes at this point.  Consequently, all
wave functions we experimented with contained the condition that
$\psi^{'}$ vanishes at $\langle r_{Z} \rangle=\alpha \langle r_{s}
\rangle$, namely contained the factor $(r-\alpha\langle r_{s}
\rangle)^{2}$.  Here $\langle r_{s} \rangle$ is the distance to the
nearest ion irrespective of its charge (in spherical packing).  Note
that in the Sun ${\rm Be}^{7}$ is a trace element and hence the
nearest neighbor would most probably be a proton or a Helium nucleus. 
The approximation can be considered as a muffin-tin potential with a
modified cell size, namely Coulomb inside $\langle r_{Z} \rangle$ and
constant outside (cf.  \citealt{Lai91}).  It is obvious that imposing the
above condition on the trial function increases the eigenvalue and
hence ionization would occur, assuming all other conditions are
unchanged, at a lower density.

Since we have a mixture of various ions, the point at which the force
between ions vanishes varies with the ion.  If our ion has a charge $Z$,
then the corresponding ionic radius $\langle r_{Z} \rangle$ is given
by:
\begin{equation}
	\langle r_{Z} \rangle = {\delta \over 1 + \delta } \langle r_{s} 
	\rangle  \quad {\rm where} \quad \delta = 
	\left( {{{Z\sum {X_iZ_i} / A_i} \over {\sum {X_i / A_i}}}} \right)^{1 / 2}.
	\label{eq:r(Z)}
\end{equation}
If all ions are equal, then $\langle r_{Z} \rangle = (1/2)\langle
r_{s} \rangle $.  Clearly, the above definition is a generalization of
the Wigner-Seitz cell idea to a mixture of species.  In the present
case, we assume the electron to be localized in the Wigner-Seitz cell. 
We will define  complete ionization when the localization of the electron
ceases. 
In the DH case, there is no such assumption.  Nevertheless, there are
other ions (and electrons) moving inside the Bohr radius.

The relation of $\langle r_{Z} \rangle$ or  $\langle r_{s} \rangle$ 
to $R_{D}$ is interesting. The condition $\langle r_{s} \rangle = 
R_{D}$ can be written as
\begin{equation}
\rho _{crit}({\rm g~cm^{-3}})=1.57(1+3X)^2\left( {{{T_6} \over {3+X}}} \right)^3.
	\label{eq:crit}
\end{equation}
For densities below the critical density, the mean interparticle
distance is smaller than the Debye radius.  Hence, when analyzing the
possibility that an ion carries a bound electron, the Debye length is
not relevant.  In fig.~\ref{fig:the-sun-rho-crit} we show the run of
the density and temperature in the Sun along with the critical density
(calculated for the actual temperature and composition).
  Fig.~\ref{fig:rho-T-plane} depicts the domains  of
$R_{D}=\langle r_{s} \rangle $ for three values of the hydrogen mass
fraction, 0.35, 0.5 and 0.7.  One finds that through most of the
volume of the Sun, the density is always below the critical one and
hence the analysis of the structure of the electronic levels must take
into account the nearby ion rather than the Debye radius.  Only close
to the surface does the situation change, the Debye radius becomes
smaller than the mean interparticle distance and the critical density
becomes smaller than the actual density.

\bfig \epsfig{file=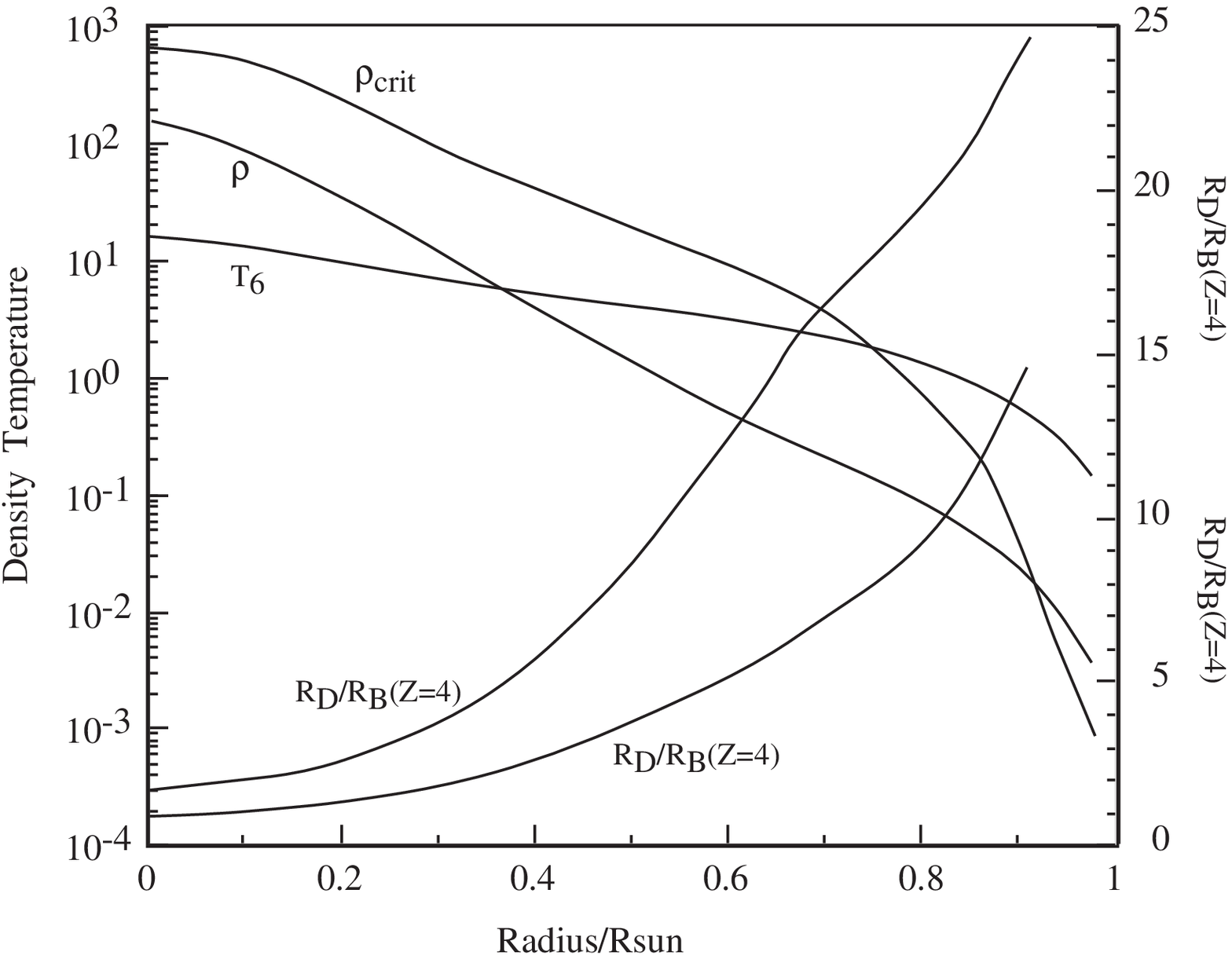,width=\figurewidth}
\caption{The run of the temperature (${\rm T}_{6}$), the density, and
the critical density as well as the Debye radius and the mean
interparticle distance (the latter given in units of the Bohr radius
for a $Z=4$ ion) in the Sun.  }
\label{fig:the-sun-rho-crit}
\efig

\bfig \epsfig{file=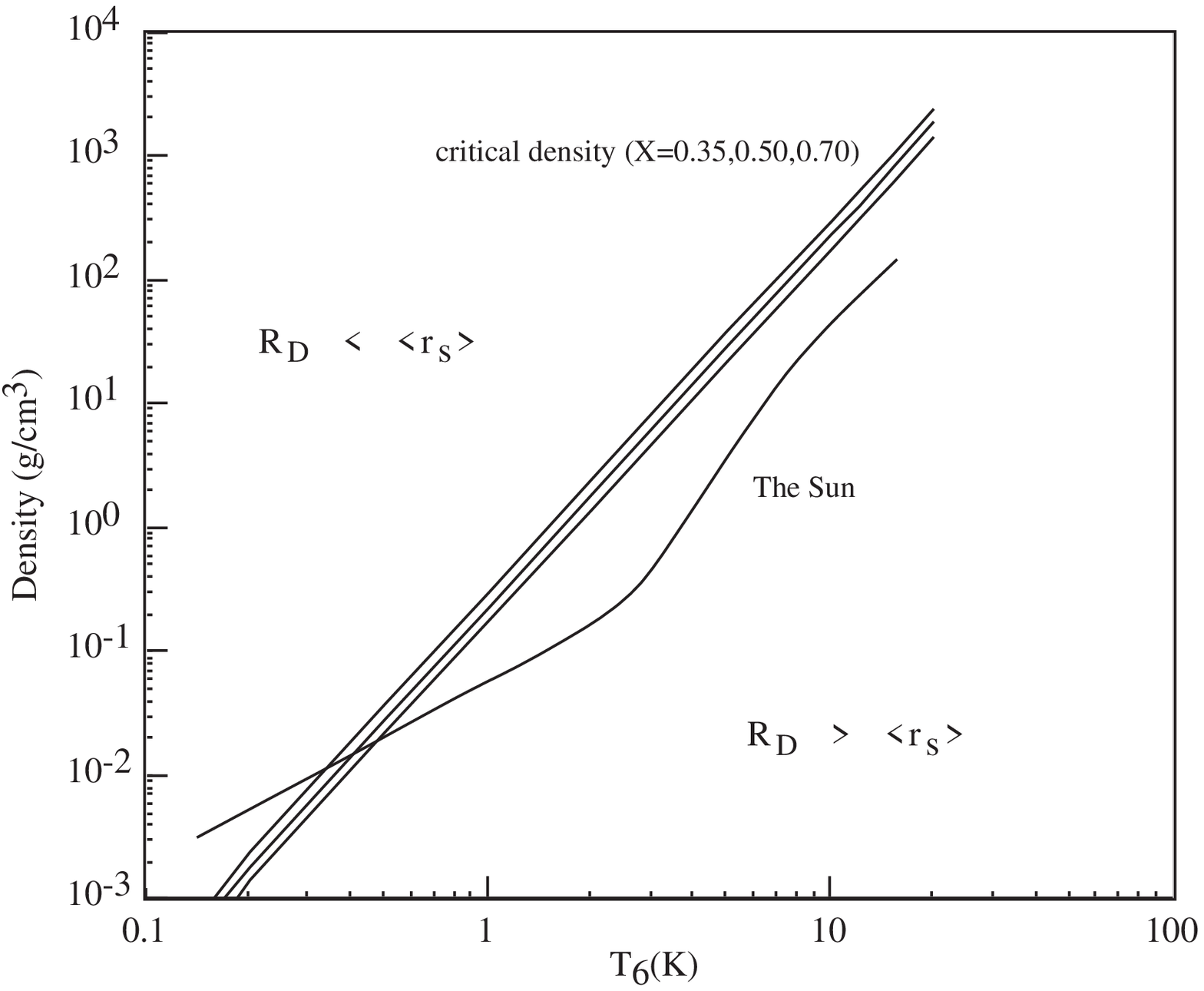,width=\figurewidth} \caption{The
temperature-density plane with the structure line of the Sun and the
lines $R_{D}=\langle r_{s} \rangle$ for three different values of the
Hydrogen mass fraction, 0.35, 0.5 and 0.7.  The almost straight part of
the structure line is the convective zone.}
\label{fig:rho-T-plane} 
\efig

When one evaluates the pressure ionization for metals at $T=0$, one
assumes the Wigner Seitz cell.  The rational for using it at higher
temperatures is the fact that the speed of the electron in the bound
state is so much greater than the speed of the ions, such that the ions
can be assumed to be at rest. The use of the ion-sphere for opacity calculations
was examined by \citet{Roz92}.

\subsection{ Schr\"{o}dinger equation with Coulomb potential }
We used the variational principle method and trial functions of two
types.  The first type is taken from \citet{Rous74}, namely a
polynomial in $r$ times an exponential function (the simple bound
$s$-state), while the second type is a Pad\'e approximation.

The results for Hydrogen obtained using the two types of trial
functions are compared and found to be practically the same to within
a relative accuracy of $10^{-2}$ or better.  Additional trials with
other functions and parameters did not improve the results beyond the
second significant digit.

Interestingly, with our definition of $ \langle r_{Z} \rangle$, the
energy level is only a function of $ \langle r_{Z} \rangle /R_{B}(Z)$
and it is shown in fig.~\ref{fig:E(hxi)}.  Complete pressure
ionization is found to occur at $ \langle r_{Z} \rangle=1.945R_{B}(Z)$
irrespective of Z. To obtain the results for a particular ion, one has
to find its $ \langle r_{Z} \rangle$ for the composition, temperature
and density under consideration.  The results for $T=0$ are shown in
fig.~\ref{fig:Schr-coul-finite-rz}.  We stress that these results are
obtained for $T=0$ and do not depend on the Debye radius (and hence do
not depend on the temperature indirectly).  We conclude that neither
${\rm Ne}^{20}$, nor species with a higher Z, are fully ionized in the
solar core.

\bfig
\epsfig{file=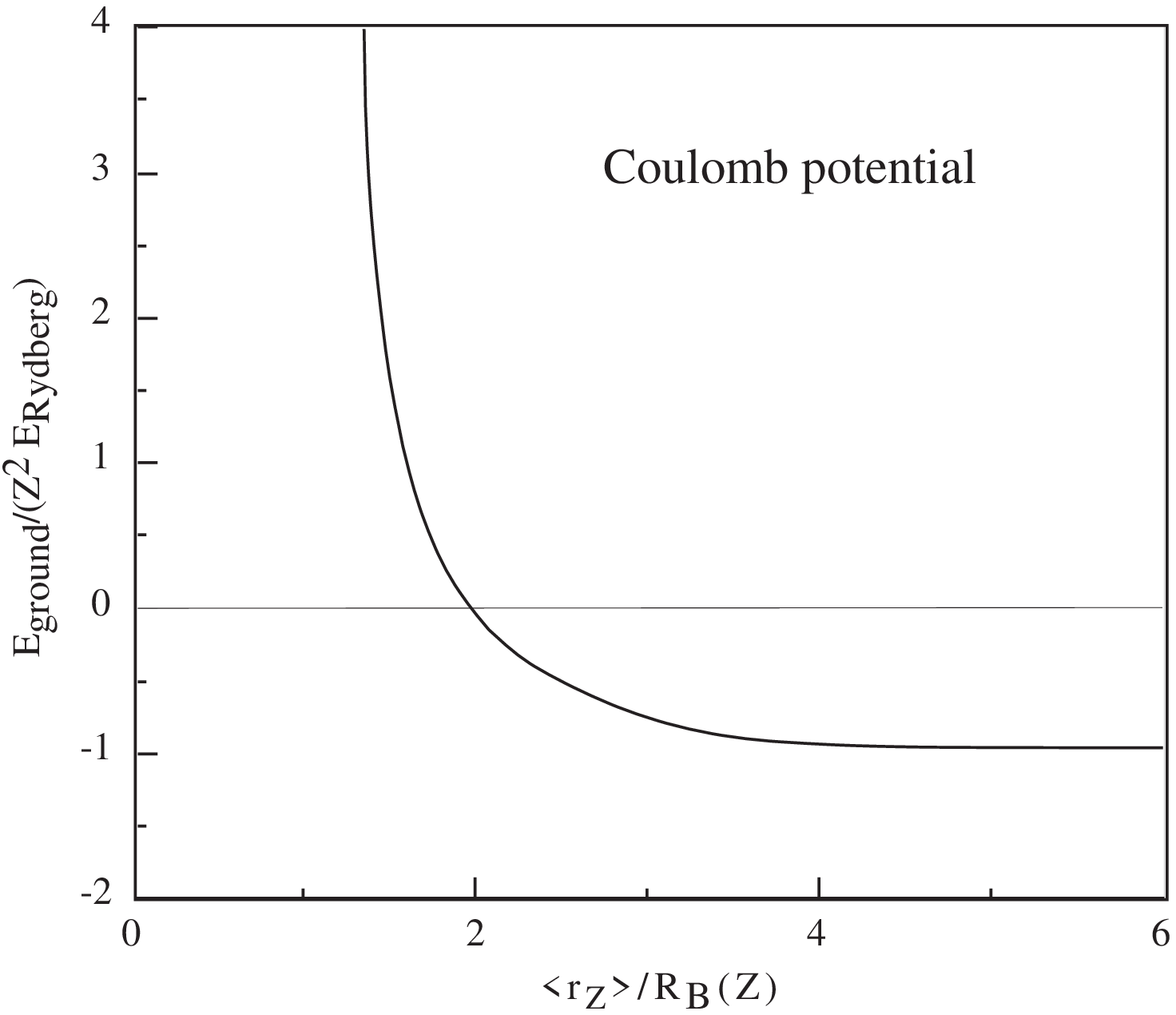,width=\figurewidth}
\caption{The energy of the ground state of an ions with charge Z 
immersed in a plasma inside a unit cell of size $\langle r_{Z} 
\rangle$.   }
\label{fig:E(hxi)}
\efig

\bfig
\epsfig{file=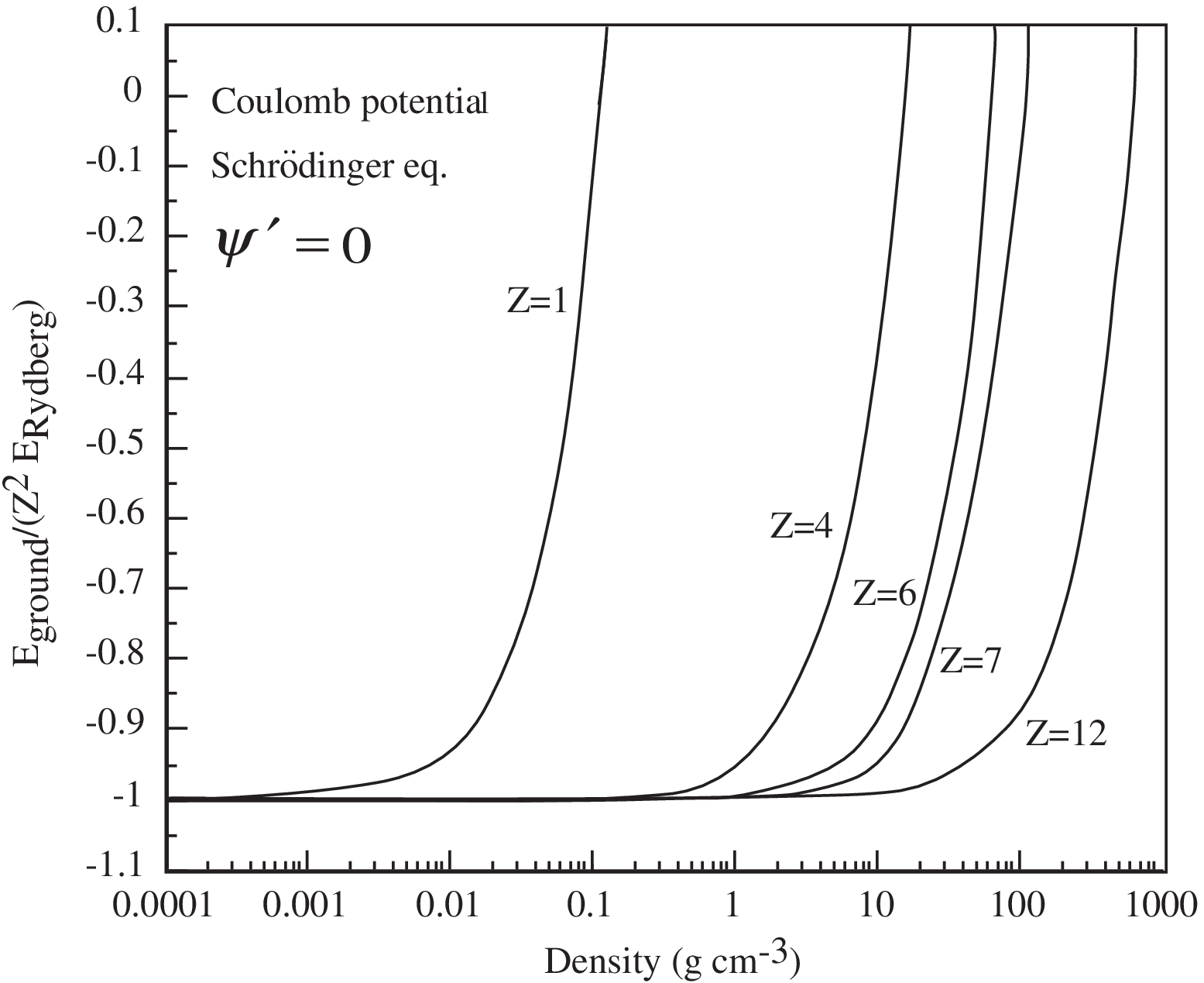,width=\figurewidth}
\caption{The translation of the previous figure to the density 
dependence of the energy levels of various species. All calculations 
assume that the $Z \ne 1$ ions are trace elements.  }
\label{fig:Schr-coul-finite-rz}
\efig

The above results are easily translated into the conditions in the Sun. 
In fig.~\ref{fig:the-sun1}, we plot the run of the ground state 
binding energy of 
${\rm Be}^{7}$ throughout the Sun. This calculation indicates that 
Beryllium is fully ionized in the Sun below a solar mass fraction of 0.66.

\bfig
\epsfig{file=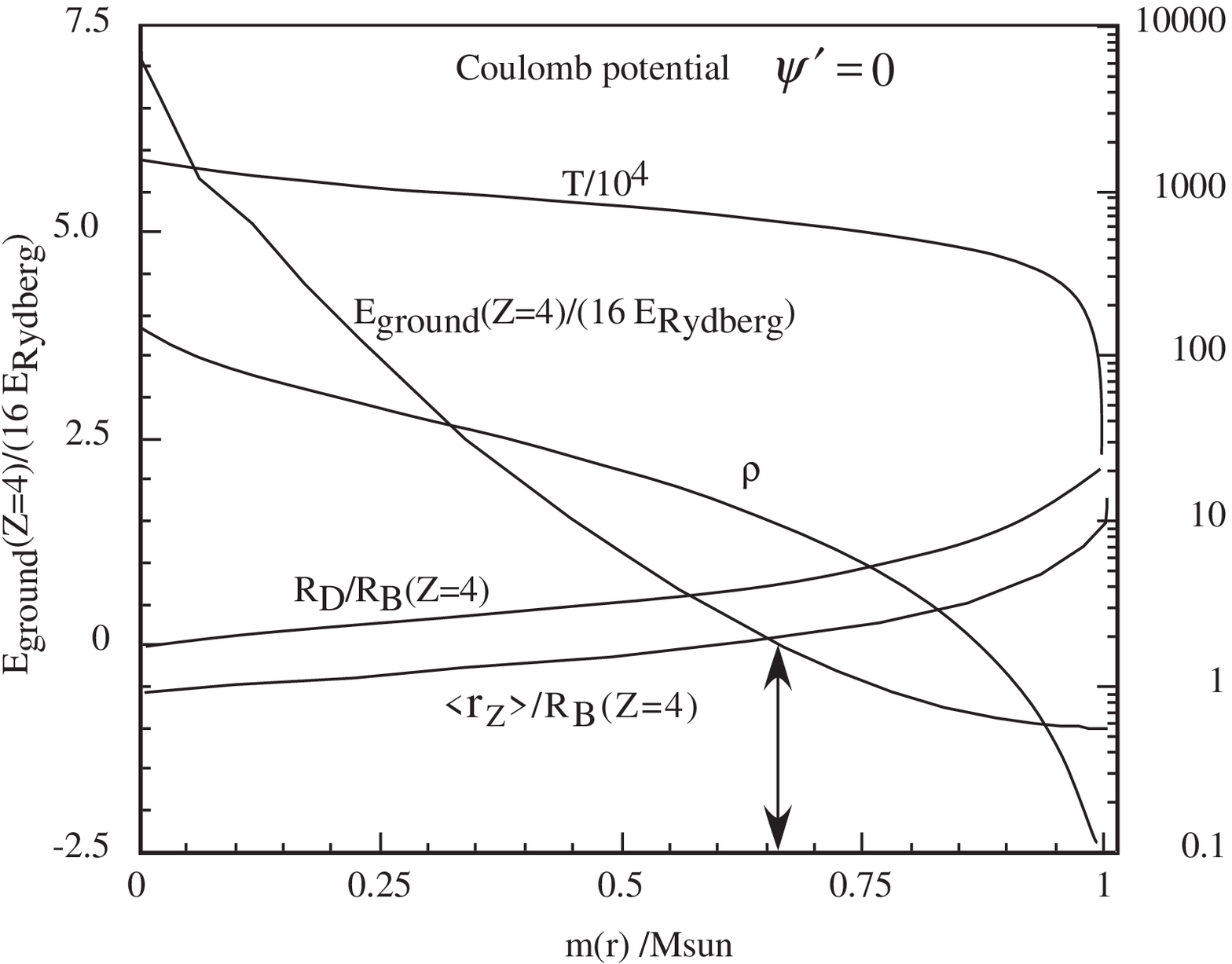,width=\figurewidth}
\caption{The run of the ground level of ${\rm Be}^{7}$ in the Sun 
along with relevant thermodynamic parameters. On the left is the ground
 energy level axis. The axis for all other quantities is on the right. Temperature is
given in $10^4{\rm K}$, density in ${\rm g~cm}^{-3}$.
The arrow marks the point of complete ionization.}
\label{fig:the-sun1}
\efig

The critical densities for the disappearance of the bound state in the
corresponding Hydrogen like ions are given in Table \ref{tbl:crit}. 
At a finite temperature, complete ionization takes places at
somewhat lower densities because the temperature increases the
excitation and with it the probability of ionization.

We conclude that the densities at which full ionization of ${\rm Be}^7$ takes place
are significantly lower than those found in the solar core. 
Thus, the small inaccuracies in the present estimate have no practical
effect on the state of Be in the solar core.  The effect on the
entropy density of the envelope is yet another issue, and it will be
discussed elsewhere.

\subsection{A Screened Potential?}
The Schr\"{o}dinger equation was solved using the simple Coulomb
potential.  It did not incorporate the Debye screened potential. 
However, electron screening does take place and cannot be ignored. 
Should the Debye screened potential be used?  Suppose that $R_{D}\gg
\langle r_{s} \rangle$ (which is not the case in the Sun), namely
there are many particles in the Debye sphere.  The Debye sphere
contains electrons and ions, hence under these conditions there are
ions which are closer to the specific ion than the Debye radius. 
These ions disturb the given ion and one should look for the bound
level under these conditions, namely that there is another ion close
by.  This is exactly what has been done above.  Thus, when the Debye
radius is very large relative to the mean interparticle distance it
creates a constant potential at the location where the two close
particles are and the effect is a shift in the energy and pressure
ionization of the very high energy levels.  On the other hand, when
$R_{D}\approx \langle r_{s} \rangle$ the Debye length and potential
lose their meaning.  This is exactly the situation in the solar core. 
A full treatment of this limit which incorporate a Debye (or a more
accurate) potential would have led to a still lower densities for the
disappearance of the bound state).

\subsection{The Kohn-Sham equation}
There are two major deficiencies in the above treatment.  The first
one was discussed at length and has to do with the doubtful validity
of the Debye potential under solar conditions and for the question of 
Be pressure ionization.  A possible way to overcome this problem is to use a
density functional (cf.  \citealt{DP82} where  it is applied to Hydrogen
plasma and where bound states are found).  The second deficiency is
the neglect of the free electrons and their effect on the screening of
the nucleus, correlations and exchange.  We therefore resorted
to the Kohn-Sham equation (\citealt{KS65}), in which these deficiencies
are taken care of, for finding the density at which pressure
ionization takes place.  The Kohn-Sham equation takes the $N$ electron
wave function and treats it as a collection of single particle
eigenfunctions.  The governing equation of the Kohn-Sham density
functional method is then:
\begin{eqnarray}
-{{\hbar^{2}} \over {2m}}\nabla ^2\psi_l({\bf r})+\left[ {-{{Ze} \over{\bf r}}
+
e^2 \int {d{\bf r'}{{n({\bf r'})} \over {\left| {{\bf r}-{\bf r'}} \right|}} }} \right. & & \\ \left. {{ -
\left( {{3 \over \pi } n({\bf r})} \right)^{1/ 3}}}\right]\psi _l(r)&=&E_l\psi _l({\bf r}). \nonumber 
	\label{eq:KS1}
\end{eqnarray}
where $n({\bf r})$ is the electron density and $\psi_{l}$ is the
single electron wave function.  The Kohn-Sham equation (hereafter KS
equation) has additional merits (\citealt{Mard00}).  The major problem
with Thomas Fermi equation is that high densities do not necessarily
lead to high kinetic energies for the electrons.  This problem is cured
in the above KS equation.  There are more merits of using the
KS equation in our particular problem.  In the Hartree-Fock approach,
the many-body wave function in form of a Slater determinant plays the
key role in the theory.  The Hartree-Fock equation if derived by
minimization of the total energy is expressed by a determinant of
wave functions which is extremely difficult to handle.  In the density
functional theory the key role is played by the observed quantity, the
electron density.  The Hohenberg-Kohn theorem then shows that for
ground states the density functional theory possess an exact energy
functional and there exists a variational principle for the electron
density.  The KS equation is then an effective one electron
equation where the exchange operator in the Hartree-Fock equation is
replaced by an exchange-correlation operator that depends only on the
electron density.  This is exactly what is needed in the present
problem.  The KS equation then treats the $N$ electron problem as single
electron wave functions.  

Let $n({\bf r})=\sum_{l=1}^N
{\left| {\psi_l({\bf r})} \right|}^2$, where the summation is carried
over all electrons, be the electron density.  In our particular case,
we examine the bound state of Be (Z=4) (as well as that of the higher Z
trace species like C, N etc).  The Be is a trace element immersed in a plasma
of fully ionized Hydrogen and Helium (mostly) and negligible amounts
of heavy elements.  Hence, the major contribution to the electron
density comes from the electrons contributed by Hydrogen and Helium. 
This term is essentially given by the environment in which the trace
specie is immersed.  Returning to the KS equation, the third term in
the KS equation is the mutual electron-electron interaction between
the bound electron and the free electrons which exist inside the
`effective orbit' and it provides the effective electron screening of
the ion.  It is easy to estimate when this term becomes important. The number
of free electrons per Bohr radius, $N_B$  is given by:
\begin{equation}
N_B \simeq \frac{6.16 \times 10^{-25}}   {Z^3}n_e
\end{equation}
This term becomes important for $N_B \sim 1$ or  
$n_e \gtrsim 1.623\times 10^{24}  Z^{3}$.
The forth term is the exchange which is given by
${{\partial { E}_{ex}} / {\partial n}}$ where $E_{ex}$ is the exchange
energy.  Because of its unique properties, the KS equation gained
popularity with physical chemists.  The accuracy of using the KS
equation for the calculation of ionization potentials in molecules is
described in \citet{CRRRP98}.  As a rule, the
results of the KS equation are more accurate than those obtained from the
 Hartree-Fock
approximation and reach the accuracy required by quantum chemists. 
The implementation of the KS equation in astrophysics of dense matter
is described for the first time by \citet{Lai91}.

We solved for the eigenvalue of the KS equation assuming the
composition of the present solar core.  We used a variational
principle with several trial functions since we are mostly interested
in the eigenvalues and not in the wave functions.  The results for ${\rm
Be}^{7}$ are shown in figure \ref{fig:Be7-ioniz} along with the
results for the Coulomb potential (and the same boundary condition). 
It appears that the correlations and exchange terms in the KS
equation contribute to the further suppression of the energy level and
complete ionization occurs a l$\grave{\rm a}$ KS at a higher
density.  Note that for sufficiently low densities, the KS
predicts significant lowering of the ground state of a {\it trace}
element relative to the continuum.  This is a consequence of the
exchange term which is mostly contributed by Hydrogen and Helium and
not by the electrons of the  trace ion under consideration.  The phenomenon does not
occur for species which are not trace elements (see later).  The
results with different trial functions vary a bit and shown are the
best (lowest) results.  The critical densities and $\langle
r_{Z}\rangle$ as obtained in the two approximations are compared in
Table \ref{tbl:crit}.  From the table we see that in the Schr${\ddot
{\rm o}}$dinger approximation $\langle r_{Z}\rangle$ is constant while
in the KS equation it varies and increases with the charge of
the ion.  A good fit for the value of the critical $\langle r_{Z}
\rangle$ over this range of charges is $r_{Z}/Z=0.451-0.026Z$.  This
expression can be easily translated into a term added to the free
energy so as to secure pressure ionization.

\bfig
\epsfig{file=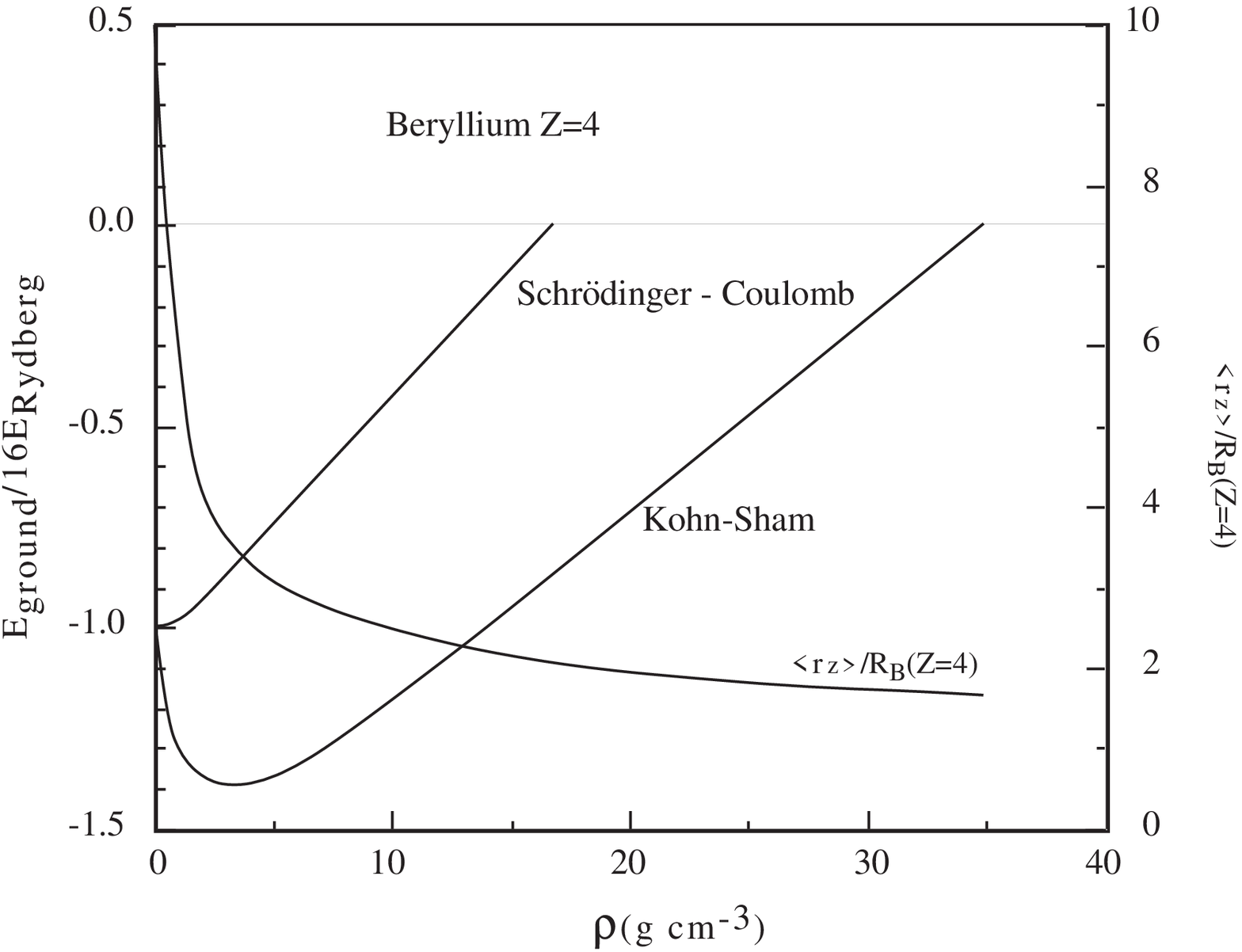,width=\figurewidth}
\caption{A comparison between the Muffin-Tin models employing the 
Schr${\ddot {\rm o}}$dinger and the Kohn-Sham equations. Also shown is 
the radius of the atomic cell $r_{Z}$ in units of the Bohr radius for a
Hydrogen like ion with $Z=4$ treated as a trace element.
The ground state energy level varies almost 
linearly with the density when close to the complete ionization density.} 
\label{fig:Be7-ioniz}
\efig

\btab
	\caption{\label{tbl:crit}The Critical Density and the radius of the atomic cell
	for the Vanishing of a Bound State in 
	Hydrogen Like Ions}
\begin{tabular}{c c c c c}
\hline
Ion & $\rho _{crit}(Coul)$ &  $\left\langle {r_Z} \right\rangle$ 
&$ \rho _{crit}(KS)$ & $ \left\langle {r_Z} \right\rangle$ \\
 Units & g cm$^{-3}$ &  $R_B(Z)$  &  g cm$^{-3}$ & $R_B(Z)$  \\
\hline
{Be$^7$}&{16.33}&{1.921}&{36.18}&{1.474}\cr
{C$^{12}$}&{65.25}&{1.923}&{86.25}&{1.752}\cr
{N$^{14}$}&{110}&{1.922}&{132.5}&{1.806}\cr
{O$^{16}$}&{171}&{1.927}&{196.5}&{1.840}\cr
{Mg$^{24}$}&{660}&{1.925}&{698.8}&{1.890}\cr
\hline
\end{tabular}
\etab

Based on the KS equation we find that the CNO elements are
fully ionized in the core of the sun.  Indeed, at $T=0$ the Oxygen still has 
a bound electron at the densities of the solar core, but the low binding 
energy and the high temperature impose complete ionization under the
 conditions in the center of the Sun. On the
 other hand, species with
$Z\ge 10$, like Neon and Iron, still keep their K-shell electrons. 
Assuming that all species heavier or equal to Ne are fully ionized
introduces a relative error of the order of $\lesssim
Z(1-X(C)-X(N)-X(O))/10$, or about $10^{-3}$ in the pressure and speed
of sound.

\subsection{The effect of the boundary condition in the Kohn-Sham equation}
The effect of the boundary condition on the result can be seen in the following way. 
We solved the KS equation under the condition $\psi(r \rightarrow \infty)=0$ for various
 densities. The result is that as the density increases monotonically, the first bound
 state becomes monotonically
more bound and pressure ionization never occurs. The free electrons are able to
 prevent ionization  exactly like the Saha equation with electron degeneracy, 
which yields 
that  the free electrons recombine as the density increases.

\subsection{Metallic Beryllium}
\citet{Per90} applied the Neutral-Pseudo Atom (NPA) method to evaluate
the equation of state and the degree of ionization of pure ${\rm Be}$
metal under normal conditions ($\rho=1.85~{\rm g~cm}^{-3}$ and $T=0$) and at
high densities, keeping $T=0$.  The results are not directly
applicable to the Sun because Perrot discusses pure metallic Beryllium
while we are interested in a trace ${\rm Be}$ atom embedded in a sea
of protons and $\alpha $ particles.  Yet, the results are instructive
for the comparison of the present method with others, especially
because they serve as a consistency check.  The applicability of the
NPA method \citep{Per90} is limited to compression ratios $c$ below 40
(see fig.~5 in the \citealt{Per90} and the explanation therein) while
Be becomes fully ionized at compression ratio of $c=50$.  At high
compressions, band calculations show that the gap
between the 1s band and the upper one close  \citep{MZ88}.  The normal density of
Beryllium is $1.85~{\rm g~cm^{-3}}$ and hence a compression ratio of 50
corresponds to a density of $92.5~{\rm g~cm^{-3}}$.  Consequently, Metallic Beryllium
is fully pressure ionized at $T=0$ and a density of $92.5~{\rm g~cm^{-3}}$.  \citet{Per90} presents an extrapolation between the end point of his
results ($c=40$) and the band calculations ($c=50$).

We note that when the density or temperature (or both) are increased starting 
from $\rho=92.5~{\rm g~cm^{-3}}$ and $T=0$  no bound state can
re-appear.  There is simply no bound state at higher densities and/or
higher temperatures if it does not exist for $T=0$ and a given
density.  


Our calculations of the pressure ionization assume  that the heavy elements
under considerations are trace elements.  Yet, it is of interest to
compare the present method with others.  The above results for
metallic Beryllium provide such an opportunity.  When the specie under
consideration is a trace element, the electrons are contributed by the
Hydrogen and Helium and there is no connection between the number of
electrons in a unit cell and the charge of the ion.  (There is no
condition of charge neutrality per each ion sphere.  The number of
electrons in the Beryllium ion sphere  is not necessarily 4).  In the pure
Beryllium case, when we search for the density of complete
ionization, we assume that the free electrons are the the first three
electrons of Beryllium and that the forth one is bound.  Thus the
exchange term is evaluated on the basis of Beryllium ionized three
times.
 

 As can be deduced from the previous calculation, a critical factor 
 is the packing of the specie. If spherical packing is assumed, then 
 the KS equation predicts  complete 
ionization at  $\rho\approx 56~{\rm g~cm}^{-3}$. On the other hand, 
\citet{Per90} quotes that complete ionization is reached at $92.5~{\rm g~ cm}^{-3}$. The exact lattice structure of Beryllium at very high 
densities is not known and Be apparently undergoes a structural change at 
a compression ratio  of about 3. However, if we assume an fcc lattice structure 
and resort to the appendix in \citet{Lai91} to find the relation between the 
lattice structure and the radius of the atomic cell 
($r_{i}=(\sqrt{2}/4)a$) we find complete ionization at a density of 
$89.3 ~{\rm g~cm}^{-3}$, in good agreement with \citet{Per90}.
This result reassures that our analysis for ${\rm Be}^7$ 
as a trace element is valid as well.

\section{The effect of ${\rm Be}$ full ionization on the solar neutrino flux}
When Beryllium is completely pressure ionized in the core of the Sun, 
 electron capture by Be takes place via the continuum only.
 Consequently, no corrections to the rate due to bound electrons
 should apply.

\btabwide
	\caption{\label{tbl:neut} The effect of treating the Be as fully ionized on the solar 
	neutrino flux}
 \begin{tabular}{c c c c c c c c c }
\hline
  & \multicolumn{2}{c}{Cas97} &  \multicolumn{2}{c}{DS96} &  \multicolumn{2}{c}{RMP98} & 
 \multicolumn{2}{c}{NACRE} \\
 & full & partial & full & partial &full & partial &full & partial  \\
\hline
${\phi (B^8) / 10^6{\rm cm^{-2}s^{-1}}}$ & 6.36 & 5.33 & 3.74 & 3.13 & 6.20 & 5.19 & 6.30 & 5.28  \\
$ {\phi (B^8)10^3 /  \phi (Be^7)}$ & 1.42 & 1.19 & 0.935 & 0.783 & 1.25 & 1.05 & 1.25 & 1.04  \\
{Ga(SNU)} & 133 & 133  &121 & 121 & 132 & 132 & 137 & 137 \\

{Cl(SNU)} & 9.14  &  7.80 & 5.90  &  5.10 & 8.96 & 7.65 & 9.29 & 7.96 \\
\hline
\end{tabular}
\etabwide

The effect of the complete ionization of ${\rm Be}^{7}$ on the solar 
neutrino flux can be easily estimated without recourse to detailed 
solar models because ${\rm Be}^{7}$ is a trace element and the amount 
of energy released by its reactions is negligible. 
Two facts are relevant. 
(a) The ${\rm Be}^{7}$ electron capture neutrino flux  
and the ${\rm B}^{8}$ flux are proportional to the 
abundance of ${\rm Be}^{7}$. (b) The rate of electron capture is much 
larger than the rate of proton capture. Consequently, a decrease of 
about 20-30\% in the rate of the ${\rm Be}^{7}+e^{-}$ capture increases the 
abundance of ${\rm Be}^{7}$ by the same factor and hence increases the
 ${\rm B}^{8}$ neutrino flux by the same factor. The flux of the ${\rm 
 Be}^{7}+e^{-}$ neutrinos is unchanged because the decrease in the rate 
 is fully compensated for by the increase in the abundance. As a matter of 
 fact, the rate of ${\rm Be}^{7}+e^{-}$ is determined (under the condition 
 that the rate of ${\rm Be}^{7}+e^{-}$ is much larger than the rate of 
 ${\rm Be}^{7}+p$ ) only by the rate of ${\rm He}^{3}+ {\rm He}^{4}$. 
The above treatment of the pressure ionization of ${\rm Be}^{7}$ 
applies to all solar models irrespective of any other parameter. The effect 
on the predicted flux assuming few different sets of nuclear parameters is 
shown in Table \ref{tbl:neut}. The four models referred to in the 
table are the following. $Cas97$ refers to \citet{Cas97}, $DS96$ 
refers to \citet{DS96}, $RMP98$ refers to \citet{RMP98} and NACRE refers 
to \citet{NACRE}. The models 
differ mainly in the parameters for the rates of the nuclear reaction 
and slightly in the details of the equation of state.

\section{Conclusion}
In this work, we were interested in the internal structure of a trace  ion in a 
high density plasma, and not in the general properties of the plasma. 
The relevant physical quantity for the latter, is the effective charge of an ion---not the 
electronic structure provided the effective radius of the electronic wave function 
is much smaller than the Debye radius.  As to the structure of the ion itself, 
we distinguished between two cases depending on whether the 
mean interparticle distance is smaller or larger than the Debye 
radius. We found that throughout most of the Sun, the mean 
interparticle distance is slightly smaller than the Debye radius.

 Under such conditions, the ionization state of a given ion 
depends mostly on the distance to the nearest neighbor rather than 
the distance scale of the screened potential. Hence, we approximate 
the condition of the plasma by a generalized Wigner-Seitz cell around 
each ion plus the proper boundary condition on the surface of the cell. 
We then  impose the Bloch 
condition on the boundary of the generalized Wigner-Seitz cell. 
The fact that the nearest neighbor is so close implies that the 
 implementation of this condition has a major effect on the point at which 
complete ionization takes place.  

We have calculated the critical density of complete ionization 
assuming at first the Schr${\ddot {\rm o}}$dinger equation and a Coulomb 
potential. Next, we implemented the Kohn-Sham equation which takes 
into account the exchange interaction due to the free electrons and 
the screening by them. Comparison of our results with the NPA 
method yields a very good agreement for the case of metallic Beryllium. 

We mention that when the boundary condition implemented in the KS equation
is $\psi(r \rightarrow \infty)=0$ the phenomenon of pressure ionization 
disappears. 

The improvements in the treatment of pressure ionization of trace elements
show that all species with $Z\leq 8$ are fully ionized in the core of the Sun. 
(The critical density for Oxygen is a bit higher than the density in the core of 
the Sun but the high temperature secure the complete ionization of Oxygen). The 
frequently applied correction to the ${\rm {Be}^{7}}$ 
electron capture rate due to bound electrons,  is not needed. 
The revised prediction for the $\phi_{\nu}¥({\rm B}^{8})$ from the 
Sun is higher by  about 20-30$\%$. 

The discussion here was centered around the high density effects. However, one should 
take into account collective effects as well (cf.~\citealt{MW98}). The energy of the electron plasma oscillations
is 
\begin{equation}
\hbar \omega_e = 3.7 \times 10^{-11} \sqrt{n_e}~eV.
\label{eq:plasma}
\end{equation}
At an electron number density of $10^{26}$ the energy of the plasma oscillations amounts to
370 eV more than the ionization potential of ${\rm  Be}^7$. Thus the bound states of ${\rm Be}^7$
have been broadened into continuum states long before. 

\section{Acknowledgment}
G.S would like to thank the Asher Space Research Fund for partial support 
of this research.

\newcommand{\apj}{ApJ}
\newcommand{\prb}{Phys.~Rev.~B.}
\newcommand{\pra}{Phys.~Rev.~A.}
\newcommand{\pre}{Phys.~Rev.~E.}
\newcommand{\jcp}{J.~Chem.~Phys.}
\bibliography{Be7-e}

\begin{thebibliography}{}

\bibitem[\protect\citeauthoryear{{Adelberger} \& {et al.}}{{Adelberger} \& {et
  al.}}{1998}]{RMP98}
{Adelberger} E.~G.,  {et al.} 1998, Reviews of Modern Physics, 70, 1265

\bibitem[\protect\citeauthoryear{{Angulo} \& {et al.}}{{Angulo} \& {et
  al.}}{1999}]{NACRE}
{Angulo} C.,  {et al.} 1999, Nuclear Physics A, 656, 3

\bibitem[\protect\citeauthoryear{{Bahcall} \& {Pinsonneault}}{{Bahcall} \&
  {Pinsonneault}}{1992}]{BP92}
{Bahcall} J.~N.,  {Pinsonneault} M.~H.,  1992, Reviews of Modern Physics, 64,
  885

\bibitem[\protect\citeauthoryear{{Castellani}, {degl'Innocenti}, {Fiorentini},
  {Lissia} \& {Ricci}}{{Castellani} et~al.}{1997}]{Cas97}
{Castellani} V.,  {degl'Innocenti} S.,  {Fiorentini} G.,  {Lissia} M.,
  {Ricci} B.,  1997, Phys.~Rep., 281, 309

\bibitem[\protect\citeauthoryear{{Chiu} \& {Ng}}{{Chiu} \& {Ng}}{1999}]{Chiu99}
{Chiu} G.,  {Ng} A.,  1999, \pre, 59, 1024

\bibitem[\protect\citeauthoryear{{Curtiss}, {Raghavachari}, {Redfern},
  {Rassolov} \& {Pople}}{{Curtiss} et~al.}{1998}]{CRRRP98}
{Curtiss} L.~A.,  {Raghavachari} K.,  {Redfern} P.~C.,  {Rassolov} V.,
  {Pople} J.~A.,  1998, \jcp, 109, 7764

\bibitem[\protect\citeauthoryear{{Dar} \& {Shaviv}}{{Dar} \&
  {Shaviv}}{1996}]{DS96}
{Dar} A.,  {Shaviv} G.,  1996, \apj, 468, 933

\bibitem[\protect\citeauthoryear{{Dharma-Wardana} \& {Perrot}}{{Dharma-Wardana}
  \& {Perrot}}{1982}]{DP82}
{Dharma-Wardana} M.~W.~C.,  {Perrot} F.~.,  1982, \pra, 26, 2096

\bibitem[\protect\citeauthoryear{{Gruzinov} \& {Bahcall}}{{Gruzinov} \&
  {Bahcall}}{1997}]{Gruz97}
{Gruzinov} A.~V.,  {Bahcall} J.~N.,  1997, \apj, 490, 437

\bibitem[\protect\citeauthoryear{{Iben}, {Kalata} \& {Schwartz}}{{Iben}
  et~al.}{1967}]{IKS67}
{Iben} I.~J.,  {Kalata} K.,    {Schwartz} J.,  1967, \apj, 150, 1001

\bibitem[\protect\citeauthoryear{{Johnson}, {Kolbe}, {Koonin} \&
  {Langanke}}{{Johnson} et~al.}{1992}]{john92}
{Johnson} C.~W.,  {Kolbe} E.,  {Koonin} S.~E.,    {Langanke} K.,  1992, \apj,
  392, 320

\bibitem[\protect\citeauthoryear{{Kohn} \& {Sham}}{{Kohn} \&
  {Sham}}{1965}]{KS65}
{Kohn} W.,  {Sham} L.~J.,  1965, Physical Review, 140, 1133

\bibitem[\protect\citeauthoryear{{Lai}, {Abrahams} \& {Shapiro}}{{Lai}
  et~al.}{1991}]{Lai91}
{Lai} D.,  {Abrahams} A.~M.,    {Shapiro} S.~L.,  1991, \apj, 377, 612

\bibitem[\protect\citeauthoryear{{Marder}}{{Marder}}{2000}]{Mard00}
{Marder} M.~P.,  2000, {Condensed Matter Physics}.
{John Wiley}

\bibitem[\protect\citeauthoryear{{Meyer-Ter-Vehn} \& {Zittel}}{{Meyer-Ter-Vehn}
  \& {Zittel}}{1988}]{MZ88}
{Meyer-Ter-Vehn} J.,  {Zittel} W.,  1988, \prb, 37, 8674

\bibitem[\protect\citeauthoryear{{Murillo} \& {Weisheit}}{{Murillo} \&
  {Weisheit}}{1998}]{MW98}
{Murillo} M.~S.,  {Weisheit} J.,  1998, \prb, 302, 1

\bibitem[\protect\citeauthoryear{{Perrot}}{{Perrot}}{1990}]{Per90}
{Perrot} F.,  1990, \pra, 42, 4871

\bibitem[\protect\citeauthoryear{{Rogers}, {Graboske} \& {Harwood}}{{Rogers}
  et~al.}{1970}]{RGH70}
{Rogers} F.~J.,  {Graboske} H.~C.,    {Harwood} D.~J.,  1970, \pra, 1, 1577

\bibitem[\protect\citeauthoryear{{Roussel} \& {O'Connell}}{{Roussel} \&
  {O'Connell}}{1974}]{Rous74}
{Roussel} K.~M.,  {O'Connell} R.~F.,  1974, \pra, 9, 52

\bibitem[\protect\citeauthoryear{{Rozsnyai}}{{Rozsnyai}}{1992}]{Roz92}
{Rozsnyai} B.~F.,  1992, \apj, 393, 409

\end{thebibliography}

 \label{lastpage}

\end{document}